\documentclass{ifacconf}

\usepackage{graphicx}      %
\usepackage{natbib}        %
\usepackage{amsmath}
\usepackage{amssymb}
\usepackage{bm}
\usepackage{mathtools}
\usepackage{siunitx}
\usepackage{glossaries}

\usepackage{algorithm}
\usepackage{algpseudocode}
\usepackage{lipsum}

\usepackage{subcaption}
\usepackage{subfig}

\usepackage[short,nocomma]{optidef}

\usepackage{tikz}
\usepackage{circuitikz}
\usepackage{graphicx}
\usetikzlibrary{external}
\usetikzlibrary{positioning} 
\usetikzlibrary{fit}
\usetikzlibrary{calc}
\usetikzlibrary{spy}
\usetikzlibrary{plotmarks}
\usetikzlibrary{arrows.meta}
\usetikzlibrary{shapes.misc}
\usepackage{pgfplots}
\pgfplotsset{compat=newest}
\usepgfplotslibrary{patchplots}

\newacronym{mpc}{MPC}{model predictive control}
\newacronym{bo}{BO}{Bayesian optimization}
\newacronym[
    longplural={lithium-ion batteries}
]{lib}{LIB}{lithium-ion battery}
\newacronym{ecm}{ECM}{equivalent circuit model}
\newacronym{soc}{SOC}{state-of-charge}
\newacronym[%
  longplural={Gaussian processes}
]{gp}{GP}{Gaussian process}
\newacronym{cc}{CC}{constant current}
\newacronym{cv}{CV}{constant voltage}

\begin{document}
\begin{frontmatter}

\title{Learning Model Predictive Control Parameters via Bayesian Optimization for Battery Fast Charging\thanksref{footnoteinfo}} 
\thanks[footnoteinfo]{© 2024 the authors. This work has been accepted to ADCHEM 2024, IFAC for publication under a Creative Commons Licence CC-BY-NC-ND.}

\author[First]{Sebastian Hirt} 
\author[First]{Andreas Höhl} 
\author[First,Second]{Joachim Schaeffer}
\author[First]{Johannes Pohlodek}
\author[Second]{Richard D. Braatz}
\author[First]{Rolf Findeisen}

\address[First]{Technical University of Darmstadt, 
   Darmstadt, Germany \{sebastian.hirt, rolf.findeisen\}@iat.tu-darmstadt.de}
\address[Second]{Massachusetts Institute of Technology, MA, USA}

\begin{abstract}
Tuning parameters in model predictive control (MPC) presents significant challenges, particularly when there is a notable discrepancy between the controller's predictions and the actual behavior of the closed-loop plant. This mismatch may stem from factors like substantial model-plant differences, limited prediction horizons that do not cover the entire time of interest, or unforeseen system disturbances. Such mismatches can jeopardize both performance and safety, including constraint satisfaction. Traditional methods address this issue by modifying the finite horizon cost function to better reflect the overall operational cost, learning parts of the prediction model from data, or implementing robust MPC strategies, which might be either computationally intensive or overly cautious.
As an alternative, directly optimizing or learning the controller parameters to enhance closed-loop performance has been proposed. We apply Bayesian optimization for efficient learning of unknown model parameters and parameterized constraint backoff terms, aiming to improve closed-loop performance of battery fast charging. This approach establishes a hierarchical control framework where Bayesian optimization directly fine-tunes closed-loop behavior towards a global and long-term objective, while MPC handles lower-level, short-term control tasks.
For lithium-ion battery fast charging, we show that the  learning approach not only ensures safe operation but also maximizes closed-loop performance. This includes maintaining the battery's operation below its maximum terminal voltage and reducing charging times, all achieved using a standard nominal MPC model with a short horizon and notable initial model-plant mismatch.
\end{abstract}

\begin{keyword}
Closed-loop Learning, Policy Optimization, Controller Autotuning, Model Predictive Control, Bayesian Optimization, Battery Fast Charging
\end{keyword}

\end{frontmatter}

\section{Introduction}
\Gls{mpc} allows for the optimal control of linear and nonlinear systems while explicitly taking constraints on the inputs, states, and outputs of the system into account. The closed-loop performance, however, depends on many factors, spanning from the quality of the prediction model used in the controller, appearing disturbances, to a suitable formulation of the cost function and constraints. Selection of these components and of suitable controller parameters is challenging \citep{lu2021bayesian}.

Methods have been proposed to enhance the performance of \gls{mpc} by exploiting machine learning  algorithms \citep{mesbah2022fusion,himmel2023machine}. Usually, this approach learns a prediction model or parts thereof, e.g., using Gaussian processes \citep{kocijan2016modelling} or neural networks~\citep{hohl2023path}. However, a good prediction model does not necessarily lead to good closed-loop performance, depending on the global control objective \citep{krishnamoorthy2023tuning,sorourifar2021data}. Thus, more recently, direct tuning of parametrized \gls{mpc} policies has been considered. Generally, all components of an \gls{mpc} formulation can be parametrized, such as the cost function, the prediction model, and the constraints. Usually, the parameters are tuned either by exploiting reinforcement learning strategies \citep{zanon2021safe} or \gls{bo} \citep{paulson2023tutorial,sorourifar2021data}, to improve the closed-loop performance. Typical closed-loop objectives include, e.g., fast convergence to a setpoint or constraint satisfaction. In this work, we consider \gls{bo} as it is a sample efficient and gradient-free method that is suitable for the optimization of black-box functions and consider it for learning \gls{mpc} parameters for fast charging of \glspl{lib} and optimal closed-loop performance. 

Learning/tuning of \gls{mpc} parameters in an outer loop results in a hierarchical  framework, where \gls{bo} offers global optimization of the closed-loop performance over longer periods or multiple episodes while the model predictive controller manages short-term planning. This allows offloading computational resources from the \gls{mpc} to the \gls{bo}, where the latter can be run online but does not need to fulfill strict real-time constraints, compared to the \gls{mpc}. Moreover, the \gls{mpc} acts as a structured safety layer, ensuring key control theory requirements like closed-loop stability, robustness, and constraint satisfaction. In this setting, it is crucial to carefully define the global (BO) and local (MPC) cost functions to achieve the desired closed-loop objective.

We demonstrate the approach to fast charging of \Glspl{lib}, which are essential for storing electrical energy and are a backbone for a carbon-neutral society. Due to high volumetric and gravimetric energy density, \glspl{lib} are the storage medium of choice for applications such as consumer electronics, and electric vehicles. However, purchase costs and range anxiety are hurdles for wider adoption of electric vehicles \citep{egbue2012barriers}. Optimal fast charging, i.e., charging batteries fast while keeping degradation low \citep{attia2020closed, matschek2023necessary}, is therefore paramount to avoid oversized battery systems which inflate electric vehicle prices and to allow long trips without excessive charging times.

Much attention has been devoted to applying machine learning to \glspl{lib} to address various systems analysis and design problems, e.g., see  \citep{schaeffer2024interpretation} and citations therein.  \glspl{lib} are nonlinear dynamical systems, and time-varying due to time- and usage-based degradation, and  are an intriguing application for \gls{mpc}, e.g., \citep{XAVIER2015374, chen2021novel, lucia2017towards, klein2011optimal}. To keep computational cost reasonable, battery management systems usually use \glspl{ecm} for control and estimation of the \gls{soc} \citep{plett2015battery}. 

Here, we apply \gls{bo} over multiple closed-loop episodes for optimization of a model predictive controller for battery charging. We explore the combination of \gls{mpc} for ensuring that the battery charging remains within the constraints, with \gls{bo} to simultaneously learn constraint backups and correct for model-plant mismatch.

The remainder of this article is structured as follows. Section \ref{sec:fundamentals} provides an introduction to \gls{mpc}, \gls{bo}, and \gls{ecm} battery models. Subsequently, we present two different case studies in Section \ref{sec:simulation}, followed by the conclusion in Section \ref{sec:conclusion}.

\section{Fundamentals}
\label{sec:fundamentals}
Consider a nonlinear discrete-time dynamical system,
\begin{equation} 
\label{eqn:discrete_system_general}
\begin{split}
    x_{k+1} &= f(x_k, u_k), \\
    y_k &= h(x_k, u_k),
\end{split}
\end{equation}
where $x_k$ are the system states, $u_k$ are the system inputs, $f(\cdot)$ are the system dynamics, $h(\cdot)$ is the mapping from states and inputs to the system outputs $y_k$, and $k \in \mathbb{N}_0$ is the discrete time index.
\subsection{Parameterized Model Predictive Control}
The objective is to steer system \eqref{eqn:discrete_system_general} to a desired state $(x_d, u_d)$ while satisfying constraints. One possible approach to do so is  \gls{mpc}, which is based on the repeated solution of a  finite-horizon optimal control problem \citep{rawlings2017model,findeisen2002introduction}. We assume that performance, stability, and repeated feasibility of  \gls{mpc} depends on  $n_p$ parameters $\theta \in \Theta \subset \mathbb{R}^{n_p}$, which might influence the cost, the model, or the constraints:
\begin{mini!}
    {\mathbf{\hat{u}}_k}{\left\{ \sum_{i=0}^{N-1} l_\theta({\hat x}_{i \mid k}, {\hat u}_{i \mid k}) + V_{f;\theta}({\hat x}_{N \mid k}) \! \right\}\label{eqn:mpc_ocp_cost}}{\label{eqn:mpc_ocp}}{}
    \addConstraint{\forall i}{\in \{0, 1, \dots, N-1\}: \notag}{}
    \addConstraint{}{\hat x_{i+1\mid k} = \hat f_\theta(\hat x_{i\mid k}, \hat u_{i\mid k})\!, \ \hat x_{0 \mid k} = x_k,}{\label{eqn:mpc_ocp_model}}
    \addConstraint{}{\hat x_{i \mid k} \in \mathcal{X}_\theta, \ \hat y_{i \mid k} \in \mathcal{Y}_\theta, \ {\hat u}_{i \mid k} \in \mathcal{U}_\theta, \ \hat x_{N \mid k} \in \mathcal{E}_\theta,}{\label{eqn:mpc_ocp_constraints}}
    \addConstraint{}{\hat y_{i \mid k} = \hat h_\theta(\hat x_{i \mid k}, \hat u_{i \mid k}).}{\label{eqn:eqn_mpc_ocp_output}}
\end{mini!}
Here, $\hat{\cdot}_{i\mid k}$ denotes the model-based $i$-step ahead prediction at time index $k$, $\hat f_\theta(\cdot)$ and $\hat h_\theta(\cdot)$ define the prediction model, $x_k$ is the measurement of the current system state, $N$ is the length of the prediction horizon, $l_\theta(\cdot)$ and $V_{f;\theta}(\cdot)$ are the stage and terminal cost functions, respectively, and \eqref{eqn:mpc_ocp_constraints}, are the state, output, input, and terminal constraints. 
Solving \eqref{eqn:mpc_ocp} yields an optimal input sequence $\mathbf{\hat{u}}_k^*=[\hat u_{0 \mid k}^*,\dots,\hat u_{N-1 \mid k}^*]$, of which the first element is applied to the system $u_k = \hat u_{0 \mid k}^*$.
Once new measurements are available, \eqref{eqn:mpc_ocp} is solved again at the next time index.
Similar to \citep{sorourifar2021data,lu2021bayesian} we exploit Bayesian optimization to optimize the parameters $\theta$ with respect to a closed-loop performance measure.

\glsreset{bo}
\subsection{Bayesian Optimization}
\Gls{bo} is a sample efficient, global optimization scheme for black-box functions,
\begin{equation}
    \theta^* = \arg \max_{\theta \in \Theta} \left\{G(\theta)\right\},
\end{equation}
where $\theta^*$ is the global maximizer of the objective function $G(\cdot)$ on the set $\Theta$.
In this work, $G(\cdot)$ is a predetermined but analytically unknown performance measure of the controlled system.
The optimization is achieved by sequentially learning a (probabilistic) surrogate model, which is trained on the observed samples of function $G(\cdot)$. As such, a dataset is built sequentially according to $\mathcal{D}_{n+1} = \mathcal{D}_{n} \cup (\theta_{n+1}, G(\theta_{n+1})), n \in \mathbb{N}_0$, with $\mathcal{D}_{0}=(\theta_0, G(\theta_0))$ and the initial parameters $\theta_0$.

\Glspl{gp} are commonly employed as surrogate models for \gls{bo} as they tend to be sample efficient \citep{rasmussen2006gaussian}.
Additionally, \glspl{gp} not only provide a best estimate (mean) but also an uncertainty measure (variance).
The \gls{gp} is fully defined by its mean function $m(\cdot)$ and covariance function $k(\cdot, \cdot)$ \citep{rasmussen2006gaussian}, 
\begin{equation}
    \label{eqn:posterior_gp}
    g(\xi) \sim \mathcal{GP}(m(\xi), k(\xi,\xi')).
\end{equation}
Assuming a prior mean and covariance function and given the training data $\mathcal{D}_{n}$, the posterior mean and variance for the test point $\theta$ are given by the inference procedure
\begin{equation}
\begin{split}
    m^+_n(\theta) &= m(\theta)+k(\theta,  \vartheta) k_{\gamma}^{-1} (\gamma-m(\vartheta)), \\
    k^+_n(\theta) &= k(\theta, \theta) - k(\theta, \vartheta) k_{\gamma}^{-1} k(\vartheta, \theta),
\end{split}
\end{equation}
with $\vartheta=[\theta_0,\dots, \theta_n]^\top$ and $k_\gamma=k(\vartheta,\vartheta)+\sigma^2\mathbb{I}$, where $\sigma^2$ is the noise variance on the training targets $\gamma = [G(\theta_0),\dots, G(\theta_n)]$.
The hyperparameters of the \gls{gp} are learned by optimizing the logarithmic marginal likelihood \citep{rasmussen2006gaussian}.

Based on the surrogate model, an acquisition function $\alpha(\theta)$ is set up, which exploits the mean and covariance of the \gls{gp} in the search for new parameter samples $\theta_n$.
Consequently, in each \gls{bo} iteration $n$, the acquisition function is optimized, yielding the new parameters for the next iteration,
\begin{equation}
     \theta_{n+1} = \arg \max_{\theta \in \Theta} \alpha_n(\theta).
\end{equation}
In this work, we employ the \textit{Expected Improvement} acquisition function  \citep{garnett2023bayesian}.

\subsection{Bayesian Optimization for Learning MPC Parameters}
For learning/autotuning the parameters of \gls{mpc}, Bayesian optimization is well suited as it is sample efficient, which reduces the number of necessary closed-loop trials.
In this work, we consider tuning of \gls{mpc} parameters $\theta$ for a closed-loop performance objective $G(\theta)$, e.g., distance to a desired state, settling time, or constraint satisfaction.
After determining an initial set of parameters $\theta_0$, a GP surrogate model is built in each optimization iteration $n$ based on the new dataset $\mathcal{D}_n$.
Then the acquisition function is used to specify the next parameter set $\theta_{n+1}$, the closed-loop system is run with parameters $\theta_{n+1}$, and the closed-loop performance $G(\theta_{n+1})$ is evaluated (see Algorithm \ref{alg:bo_autotuning}).
An advantage of the \gls{bo}-based \gls{mpc} tuning is the high sample efficiency and the capability to run the \gls{bo} online, thus allowing for an adaptive behavior of the \gls{mpc} to a changing system or environment.

\begin{algorithm}
\caption{MPC Parameter Learning via Bayesian Optimization}
\label{alg:bo_autotuning}
\begin{algorithmic}
\Require Parametrized MPC policy $u_0^*(x; \theta)$, Parameter domain $\Theta$, initial data $\mathcal{D}_0$, prior GP mean and covariance functions $m(\cdot)$ and $k(\cdot)$
\For{$n = 0, 1, 2, \cdots$}
    \State Determine posterior GP for $\mathcal{D}_n$
    \State Maximize acquisition function for $\theta_{n+1}$
    \State Run closed-loop simulation using $u_0^*(x; \theta_{n+1})$
    \State Determine closed-loop performance $G(\theta_{n+1})$
    \State Update data set
\EndFor
\end{algorithmic}
\end{algorithm}

\glsreset{lib}
\glsreset{ecm}
\subsection{Battery Model and Fast Charging}
\Glspl{lib} are time-varying nonlinear dynamical systems that have been modeled by empirical descriptions (e.g., lookup tables), black-box machine learning (e.g., neural networks), first-principles  (e.g., porous electrode theory models), and \glspl{ecm}  \citep{ml_and_more2023}. \Glspl{ecm} approximate the battery behavior by an electric circuit using elements such as voltage sources, resistors, capacitors, and more (e.g., \cite{krewer2018dynamic, schaeffer2023machine} and citations therein). An advantage of \glspl{ecm} over first-principle models is their low computational cost and straightforward implementation, making them suitable for control applications.

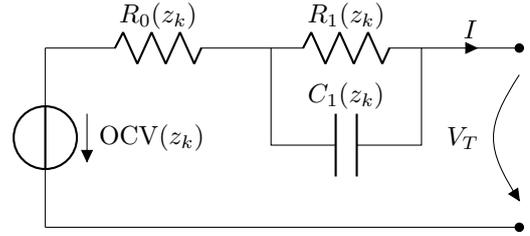
\begin{figure}[!htb]
    \centering
    \scalebox{.99}{
    \begin{circuitikz}%
        \draw (0,0) to [voltage source, v=${\text{OCV} (z_k)}$] (0,-2.4);
        \draw (0,0) to [resistor, l=$R_0 (z_k)$] (3,0)
                    to [resistor, l=$R_1 (z_k)$] (5,0);
        \draw (3,0) -- (3,-1.3)
                    to [C, l=$C_1 (z_k)$] (5,-1.3)
                    -- (5,0) to[short, i=$I$, -*] (6.3,0);
        \draw (0,-2.4) to[short, -*] (6.3,-2.4);
        \draw (6.3,0) to[open, v=$V_T$] (6.3,-2.4);
    \end{circuitikz}
    }
    \caption{R-RC battery \gls{ecm} model with parameters depending on the \gls{soc} $z_k$.}
    \label{fig:ecm}
\end{figure}

The subsequent fast charging case studies use an \gls{ecm} model with a resistor in series with a resistor capacitance pair (Fig.\,\ref{fig:ecm}). The voltage source is given by the \gls{soc}-dependent cell characteristic open circuit voltage (OCV), $R_0$ represents the ohmic resistance of the battery, and the $R_1$, $C_1$ pair models the polarization of the cell due to the double layer and approximates diffusion.

Consequently, the nonlinear discrete-time dynamics are given by
\renewcommand{\arraystretch}{1.4}
\begin{equation} 
\label{eqn:discrete_system}
\begin{split}
    \begin{bmatrix}
        z_{k+1} \\
        U_{1, k+1}
    \end{bmatrix} &=
    \begin{bmatrix}
        z_k + \frac{\eta T_s}{Q} I_k\\
        (U_{1,k}-R_1 I_k)\exp\!{\big(\!\!-\!\frac{T_s}{R_1 C_1}\big)} \!+ R_1 I_k
    \end{bmatrix}, \\
    V_{T, k} &= \mathrm{OCV} - U_{1,k} - R_0 I_k,
\end{split}
\end{equation}
where the parameters $R_0(\cdot)$, $R_1(\cdot)$, $C_1(\cdot)$, and the $\mathrm{OCV}(\cdot)$ are functions of the \gls{soc}, $U_1$ is the voltage over $C_1$, $T_s$ is the sampling time, $\eta$ is the Coulombic efficiency, and $Q$ denotes the battery discharge capacity.

We used cubic splines to interpolate the parameters published by \cite{tran2021comparative}, associated with a Samsung SDI $INR18650-20S$ lithium nickel manganese cobalt oxide (NMC) cell, and assume ideal cell cooling, i.e., constant temperature, and that the model parameters are independent of the current. As a consequence, the model parameters, i.e., $R_0$, $R_1$, and $C_1$, are only dependent on the \gls{soc} $z_k$, where $k$ is the time index. 
Batteries change their behavior not only with \gls{soc} but also with current and temperature \citep{huad_offer2021, batteries9020101}. Consequently, the error between the \gls{ecm} and the battery increases for operational points away from the parametrization region. In addition, batteries degrade over time and usage, i.e., the capacity and power capabilities decrease. Degradation does not only impact the performance of the application but also control. It furthermore changes the system dynamics, thus leading to a mismatch between model parameters and the cell, introducing a further source of model error. 

Fast charging is important to increase the utility of electric vehicles. One approach for designing fast charging protocols is to minimize the charging time subject to degradation constraints, e.g., \citep{matschek2023necessary}. Commonly, constraints are set for current, voltage, temperature, and lithium-plating overpotential to limit degradation. Here we employ current and voltage constraints given by
\begin{align}
    \forall k \in \mathbb{N}_0:\; &I_k \leq I_{\mathrm{max}} = \SI{6}{\ampere} \label{eq:imin},\\ 
    \forall k \in \mathbb{N}_0:\;\SI{2.5}{\volt}=V_{T,\mathrm{min}} \leq \;&V_{T,k} \leq V_{T,\mathrm{max}} = \SI{4.2}{\volt}\label{eq:vtmin},
\end{align}
with no temperature constraint due to the isothermal assumption. The plating overpotential constraint cannot be set because the chosen \gls{ecm} does not include plating. Note that we consider current control during the entire charging process due to the model-based approach and do not switch to voltage control as implemented in typical charging circuits, e.g., \citep{chen_fast_charging_hardware_modes}.

Next, we explore how \gls{mpc} and \gls{bo} can ensure that the voltage constraint is not violated during fast charging under model-plant mismatch. This article focuses on methodology and its demonstration, justifying the simplifications and assumptions mentioned previously.

\section{Simulation Studies}
\label{sec:simulation}
The capabilities of the \gls{bo}-\gls{mpc} framework are demonstrated in two case studies focusing on enhancing the closed-loop performance during fast charging while satisfying terminal voltage constraints. In the first case study, the \gls{bo} optimizes the voltage constraint with a constraint backoff term. In the second case study, the \gls{bo} optimizes the parameters of the controller's prediction model.

For both case studies, the fast charging problem is formulated as a set point change, where the target state is the fully charged battery, i.e., $z_k=1$. 
The \gls{mpc} is given by
\begin{mini!}
    {\mathbf{\hat{I}}_k}{\left\{ \sum_{i=0}^{N} \left(1-z_{i\mid k}\right)^2 \right\}}{}{}
    \addConstraint{\forall i}{\in \{0, \dots, N-1\}: \notag}{}
    \addConstraint{}{\text{Equation} \ \eqref{eqn:discrete_system}\notag}{}
    \addConstraint{}{V_{T,\mathrm{min}}\leq\hat V_{T, i \mid k} \leq V_{T,\mathrm{max}}.}{\label{eqn:mpc_ocp_output_2}}
\end{mini!}
We implement the state constraints as soft constraints for numerical reasons.
For the sake of simplicity, we employ the ECM model for the controller-intern predictions as well as for the closed-loop simulations. To realize the model-plant mismatch, both case studies use randomly disturbed model parameters in the controller's prediction model. The resulting model parameters differ up to $50\%$ from the true model parameters.
The nominal \gls{mpc} shows a bang-and-ride behavior, i.e., first the current is equal to its upper constraint (bang), but once the system hits the upper voltage constraint, the \gls{mpc} subsequently continues to \textit{``ride"} the upper voltage constraint. 
To avoid constraint violations with a mismatch while fast charging, the global \gls{bo} objective function is defined as
\begin{equation}
    G(\theta) = \sum_{k=0}^{M} - c_1 (1-z_k)^2 - \big(\max \{ 0, V_{T, k}-V_{T, \max}\}\big)^2,\label{eqn:bo_cost}
\end{equation}
where $c_1\ll 1$ balances the trade-off between minimization of the charging time and constraint satisfaction. Note that the closed-loop values $z_k$ and $V_{T, k}$ in \eqref{eqn:bo_cost} implicitly depend on the parameters $\theta$ of the \gls{mpc}.

\subsection{Learning Constraint Backoff}
During battery charging, it is essential not to violate the terminal voltage constraint associated with the battery chemistry and given in the battery datasheet. 
To ensure that the voltage constraints are not violated, a parametrized constraint backoff term is introduced in the \gls{mpc} via \eqref{eqn:mpc_ocp_output_2}. The resulting output constraint is
\begin{equation}
    V_{T,\mathrm{min}}\leq\hat V_{T, i \mid k} \leq V_{T,\mathrm{max}} - b_\theta(z_k),
\end{equation}
where $b_\theta(\cdot)$ is a backoff function that depends on the current \gls{soc}.
For efficient implementation, $b_\theta(\cdot)$ is defined by a cubic spline interpolation over a predefined grid. The values at the grid points are stored in $\theta \in \mathbb{R}^7$, where $7$ is the chosen number of grid points.
The \gls{bo} adjusts $\theta$ to avoid constraint violation while maximizing the state of charge, see \eqref{eqn:bo_cost}. 
The objective is to adjust $\theta$ such that the constraint is satisfied after the \gls{bo} tuning procedure.

\begin{figure}
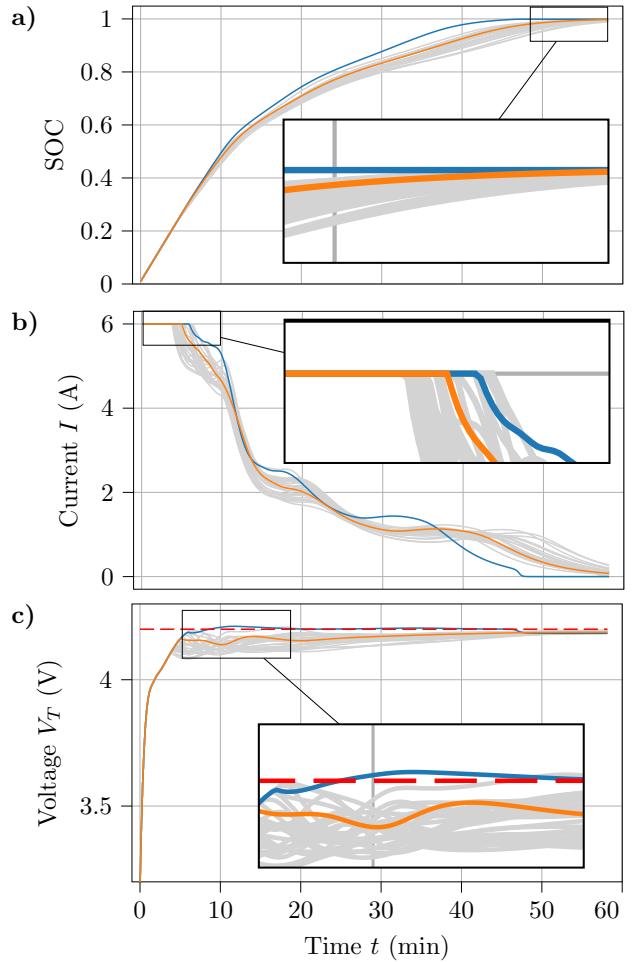

    \scalebox{0.95}{\input{figures/backoff_tuning_soc}}
    \scalebox{0.95}{\input{figures/backoff_tuning_I}}
    \scalebox{0.95}{\input{figures/backoff_tuning_Vt}}
    \vspace{-0.4cm}
    \caption{First case study: Learning constraint backoff. Initial closed-loop solution in blue, \gls{bo} trials in gray, and optimized closed-loop result in orange. \textbf{a)} SOC trajectories, \textbf{b)} Current trajectories, \textbf{c)} Voltage trajectories with voltage constraint, $V_{T, \text{max}}$, red dashes.}
    \label{fig:backoff_tuning_result}
\end{figure}

\begin{figure}
    \centering
    \input{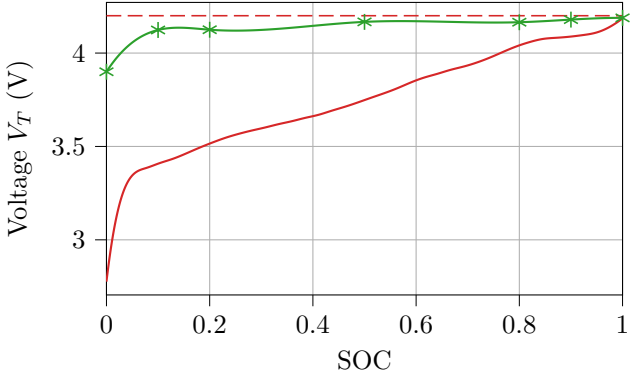}
    \vspace{-4mm}
    \caption{Spline interpolation and tuned grid points of the backed-off constraint (green) and OCV (red).}
    \label{fig:backoff_spline}
\end{figure}

The case study shows that the initial closed-loop solution violates the voltage constraint due to the model-plant mismatch (blue line, Fig.\,\ref{fig:backoff_tuning_result}c). The final voltage trajectory solution does not violate the voltage constraint because of the learned constraint backoff (orange line, Fig.\,\ref{fig:backoff_tuning_result}c). During the \gls{bo} procedure, different parameters for the constraint backoff are sampled in a structured way, where the acquisition function balances exploration and exploitation (gray lines, Fig.\,\ref{fig:backoff_tuning_result}abc).
The \gls{soc} trajectory of the optimized closed-loop result rises slower than the initial solution (blue and orange lines in Fig.\,\ref{fig:backoff_tuning_result}a).
While the initial solution promises a faster charge, this is only due to the voltage constraint violation, which results in a later switching time from \gls{cc} to \gls{cv}.
The learned constraint tightening shows that the backoff term decreases with increasing \gls{soc} (Fig.\, \ref{fig:backoff_spline}).
The high backoff at the beginning of the charging process ensures that the voltage constraint is not violated at the end of the \gls{cc} phase. %
For higher \gls{soc} values, however, the constraint backoff is smaller; otherwise, the closed-loop behavior would be overly conservative, and the battery could not be fully charged to $\mathrm{SOC} = 1$. Batteries can only charge if $\textrm{OCV} < V_T$ holds, i.e., the maximum achievable \gls{soc} is defined by the intersection of the green and red line in Fig. \ref{fig:backoff_spline}. 

As this result is still slightly conservative compared to the initial run, the next case study explores the effect of learning the prediction model parameters.

\subsection{Learning Model Parameters}
The second case study takes a different approach by learning the parameters of the controller's internal \gls{ecm}. The \gls{bo} tunes the values at the grid points of the $R_1$ spline. To demonstrate that the prediction model does not have to match the plant model exactly for a good closed-loop performance, we only tune the $R_1$ spline while keeping the mismatch for the other parameters. Note that in general, all of the \gls{ecm} parameters can be tuned.

\begin{figure}
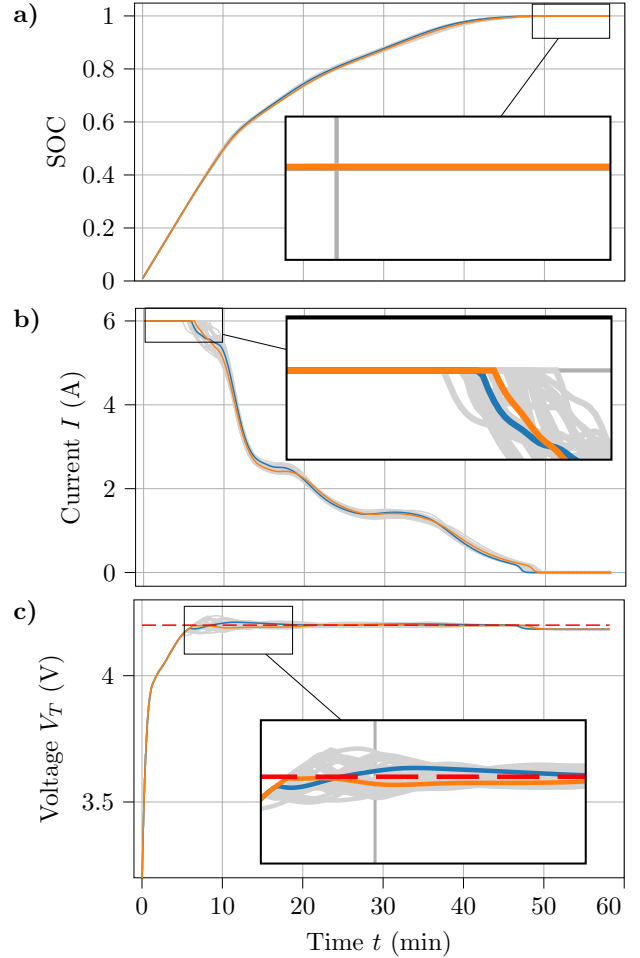

    \scalebox{0.95}{\input{figures/model_tuning_soc}}
    \scalebox{0.95}{\input{figures/model_tuning_I}}
    \scalebox{0.95}{\input{figures/model_tuning_Vt}}
    \vspace{-0.4cm}
    \caption{Second case study: Learning prediction model parameters. Initial closed-loop solution in blue, \gls{bo} trials in gray, and optimized closed-loop result in orange. \textbf{a)} SOC trajectories, \textbf{b)} Current trajectories, \textbf{c)} Voltage trajectories with voltage constraint, $V_{T, \text{max}}$, red dashes.}
    \label{fig:model_tuning_result}
\end{figure}

The terminal voltage constraint is fulfilled with the tuned system (orange line, Fig.\,\ref{fig:model_tuning_result}c). The initial trial here is identical to the initial trial of the first case study.
By tuning the prediction model directly, this case study yields a better closed-loop result, i.e., the maximum \gls{soc} is achieved about 10 minutes earlier (Fig.\, \ref{fig:model_tuning_result}a). Additionally, the charging time of the safe tuned result is only slightly lower compared to the unsafe initial result. This is also visible in the tuned current trajectory (orange line, Fig.\, \ref{fig:model_tuning_result}b), which shows a later switching time from \gls{cc} to \gls{cv} compared to the initial trajectory.
However, the trajectories of the \gls{bo} trials violate the voltage constraint as there is no explicit upper bound. 
In contrast, more \gls{bo} trials keep the voltage constraint in the first case study.
Thus, we trade off performance of the final solution with less safe sampled parameter sets.

\begin{figure}
    \centering
    \scalebox{0.95}{\input{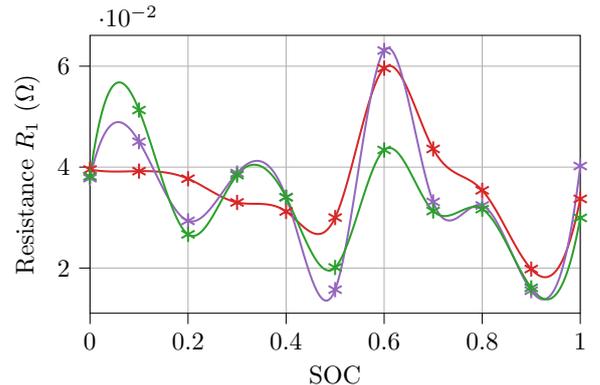}}
    \vspace{-4mm}
    \caption{Spline interpolation and tuned grid points for $R_1$ in the prediction \gls{ecm}. Plant parameter (red), initial prediction parameter (purple), and tuned prediction parameter (green).}
    \label{fig:R1_spline}
\end{figure}

In summary, the closed-loop behavior significantly improves, despite the fact that the learned model does not converge to the true model (Fig.\,\ref{fig:R1_spline}).
For $\mathrm{SOC} \approx 0.5$, the learned grid points for $R_1$ (green) are higher than for the initial parameter (purple). This is exactly where the constraint violation occurs at 12 minutes in the initial simulation (Fig.\, \ref{fig:model_tuning_result}).
A local backoff is learned through the model, making our results interpretable.
For different \gls{soc}, the learned grid points might be less intuitive to understand, as they also might be tuned towards compensating the mismatch introduced through the other parameters $R_0$ and $C_1$.


\section{Conclusion}
\label{sec:conclusion}
We exploit a hierarchical control framework  based on Bayesian optimization for global optimization of a parametrized \gls{mpc} formulation.
While the former takes care of global optimization of the long-term closed-loop behavior, the latter handles lower-level and short-time control tasks.
Two specific formulations are proposed within this framework,  which are demonstrated in case studies for fast charging of lithium-ion batteries.
The first case study learns a constraint backoff term that was parametrized over the \gls{soc} of the battery.
While mostly safe Bayesian optimization trials were observed, the final result turned out to be conservative compared to the initial unsafe trial.
The second case study had less conservative results while showing very similar performance to the initial unsafe trial.
While the final result satisfied the constraint, more trials with constraint violations were conducted by the Bayesian optimization in the second case study, indicating more exploration in the parameter space.

\bibliography{ifacconf}

\end{document}